\documentclass[fleqn,usenatbib]{mnras}

\usepackage{newtxtext,newtxmath}
\usepackage{url,times,hyperref,graphicx,color,epsfig,epstopdf}
\usepackage{cleveref}
\crefname{section}{\S}{§§}

\usepackage[T1]{fontenc}
\usepackage{ae,aecompl}

\def\head{
 \vbox to 0pt{\vss
                   \hbox to 0pt{\hskip 440pt\rm LA-UR-10-07069\hss}
                  \vskip 25pt}}

\usepackage{graphicx}	        
\usepackage{amsmath}	        
\usepackage{amssymb}	        
\usepackage{gensymb}            
\usepackage{multirow}           
\usepackage{ulem}              
\usepackage{enumitem}

\usepackage[dvipsnames]{xcolor} 

\newcommand{\Fig}[1]{Fig.~\ref{#1}}
\newcommand{\Reff}{R$_e$}
\newcommand{\SB}{$\mu_e$ }
\newcommand{\sur}{$\mu$ }
\newcommand{\mhalo}{M$_{\rm halo}$}
\newcommand{\mstar}{M$_{\star}$}
\newcommand{\msun}{M$_{\odot}$}
\newcommand{\mHI}{M$_{\rm HI}$}
\newcommand{\beq}{\begin{equation}}
\newcommand{\eeq}{\end{equation}}
\newcommand{\Sec}[1]{Section~\ref{#1}}

\title[Formation of LSBs]{NIHAO XXI: The emergence of Low Surface Brightness galaxies}

\author[Di Cintio et al.]
    {Arianna Di Cintio$^{1,2}$\thanks{E-mail: adicintio@iac.es}\thanks
    {Marie-Sk\l{}odowska-Curie Fellow}, Chris B. Brook$^{1,2}$, Andrea V. Macci\`{o}$^{3,4}$,\vspace{-.5cm} Aaron A. Dutton$^{3}$   \newauthor  \& Salvador Cardona-Barrero$^{2}$\\
\vspace{-.2cm}
$^{1}$Instituto de Astrof\'{i}sica de Canarias, Calle Via L\'{a}ctea s/n, E-38206 La Laguna, Tenerife, Spain\\\vspace{-.2cm}
$^{2}$Universidad de La Laguna. Avda. Astrof\'{i}sico Fco. S\'{a}nchez, La Laguna, Tenerife, Spain\\\vspace{-.2cm}
$^{3}$New York University Abu Dhabi, PO Box 129188, Saadiyat Island, Abu-Dhabi, United Arab Emirates\\\vspace{-.2cm}
$^{4}$Max Planck Institute f\"{u}r Astronomie, K\"{o}nigstuhl 17, D-69117, Heidelberg, Germany\vspace{-.2cm}
}

\date{Accepted XXX. Received YYY; in original form ZZZ}

\pubyear{2018}

\begin{document}
\setlength{\abovedisplayskip}{3pt}
\setlength{\belowdisplayskip}{3pt}

\label{firstpage}
\pagerange{\pageref{firstpage}--\pageref{lastpage}}
\maketitle

\begin{abstract}
The existence of galaxies with a surface brightness  $\mu$ lower than the night sky has been known since three decades. Yet, their formation mechanism and emergence within a $\rm \Lambda CDM$ universe has remained largely undetermined.
For the first time, we investigated the origin of  Low Surface Brightness (LSB) galaxies with \mstar$\sim$10$^{9.5-10}$\msun, which we are able to reproduce within hydrodynamical cosmological simulations from the NIHAO suite.
The simulated and observed LSBs share similar properties, having large HI reservoir, extended star formation histories and effective radii, low S\'{e}rsic index and  slowly rising rotation curves. 
The formation mechanism of these objects is explored: simulated LSBs  form as a result of co-planar co-rotating mergers and aligned accretion of gas at early times, while perpendicular mergers and mis-aligned gas accretion  result in higher $\mu$ galaxies by $z$=0. The larger the merger, the stronger the correlation between merger orbital configuration and final $\mu$. While the halo spin parameter is consistently high in simulated LSB galaxies, the impact of halo concentration, feedback-driven gas outflows and merger time  only plays a minor-to-no role in determining  $\mu$. Interestingly,  the formation scenario of such `classical'  LSBs  differs from the one of  less massive, \mstar$\sim$10$^{7-9}$\msun, Ultra-Diffuse Galaxies, the latter resulting from  the effects of SNae driven gas outflows: a \mstar\ of $\sim$10$^9$\msun\ thus represents the transition regime between a \textit{feedback dominated} to an \textit{angular momentum dominated}
formation scenario in the LSB realm.
Observational predictions are offered regarding spatially resolved star formation rates through LSB discs: these, together with upcoming surveys, can be used to verify the proposed emergence scenario of LSB galaxies.
 
\end{abstract}

\begin{keywords}
  methods: galaxies: haloes -- galaxies: evolution -- cosmology: theory -- dark matter
\end{keywords}


\section{Introduction} \label{sec:intro}

Low Surface Brightness galaxies (LSBs) are diffuse, faint galaxies hardly distinguishable from the night sky, whose discovery dates back to the late 80's \citep{Bothun87,impey88,schombert88}.
Although a few giant LSBs have been discovered in the field, Malin 1 being the most famous example of this category  \citep[e.g.][]{Bothun87,lelli10}, most LSB objects are found to be dwarf galaxies, with a
\mstar$\leq$10$^{10}$\msun\ (see \citealt{Impey97,Bothun97} for a review of `classical' LSBs).

LSBs do not generally have a central bulge, like regular spirals, they are dark matter (DM) dominated and show a slowly rising rotation curve \citep[e.g.][]{Dalcanton97,deBlok01,McGaugh01}, an extended star formation history (SFH) with a high fraction of young stars,  suggesting that they follow  a
similar evolutionary history as higher surface brightness galaxies,
but  at a slower rate \citep{McGaugh94,vandenHoek00,Bell00}. Most of their baryonic matter is in the form of HI gas, having amongst  the highest gas mass fractions of any studied galaxy type  \citep{Schombert01}. 
Their central ($\mu_0$) and effective ($\mu_e$) surface brightness reach much lower values than regular High Surface Brightness (HSB) objects: although a strict definition of LSBs does not exist,  the lower  limit of $\mu_e$$\geq$22-22.5 mag/arcsec$^2$ is a reasonable operative value to discriminate  these elusive galaxies from regular HSBs.
Despite representing a significant component of the galaxy population, spanning a wide range of morphological types within different  environments, their low \SB makes them hard to detect.
Only in the last decade have advances in technologies and instrumentation allowed the limit of discoveries of LSBs to be pushed to lower and lower magnitudes \citep{merritt14,vandokkum15,Fliri16,Trujillo16,delgado10}, unveiling a  large population of LSB galaxies and opening a new window on their evolution and formation. 

Further reinvigorating the interest for the LSB universe, is the recent discovery of a ubiquitous population of faint, ultra-LSB galaxies: initially identified within the Coma cluster \citep{vandokkum15,koda15,beasley16b}, they have  since then been found  in the Virgo, Fornax and other low-z clusters  \citep{mihos15,munoz15,vanderburg16,mancera18}, as well as in the field and groups \citep{delgado16,roman16,Trujillo17}. These objects, having
the stellar mass 
of small dwarf galaxies, \mstar$\sim$10$^{6.5-9}$\msun, but the effective radius, \Reff, of large
spirals, have been named Ultra-Diffuse Galaxies (UDGs), and their \SB can be as low as 28-30 mag/arcsec$^2$.

Initially referred to as a different category compared to `classical' LSBs (see, however, the 20 years old definition of Very Low Surface Brightness objects, as compared to UDGs, in \citealt{mcgaugh96}), the similarity between the two galaxy types has been recently recognized.
Indeed, UDGs exists in various colours and environments, and, just like LSBs,  can be extremely HI gas rich \citep[e.g.][]{Leisman17,Spekkens18}.

While there is not a clear consensus about the formation mechanism of UDGs, the two  current leading scenarios are: i) formation through  feedback-driven gas outflows, that are able to cause fluctuation in the potential at the center of the galaxy which, in turn, allows for an expanded and extended DM and stellar distribution \citep{dicintio17,chan18,Jiang18} and ii) formation within the high-spin tail of DM haloes \citep{amorisco16}. More recently, UDGs formation through galaxy interactions has been observationally suggested for UDGs that are found in dense environemnts \citep{Bennet18}, highlighting how different channels  can produce UDGs (see also \citealt{Chilingarian19,Carleton19}). Moreover, \citet{Yozin15} showed that  early accretion of faint galaxies into  overdense regions can further quench star formation, producing UDGs.

Within a cosmological $\Lambda$CDM context, the studies of   
 \citet{dicintio17,chan18}, based respectively on the NIHAO \citep{Wang15} and FIRE  \citep{Hopkins14}
simulation suites, have been shown to be able to reproduce a realistic population of UDGs, without any fine-tuning of the simulation parameters themselves. However, no simulations have been explicitly shown, so far, to be able to reproduce a population of `classical' LSBs, with \mstar$\sim$10$^{9-10}$\msun\ (we note here the recent work of \citealt{Zhu18}, that deals with a different, less common  type of giant-LSB, which form through cooling of hot halo gas within their simulations).

At today, the formation mechanism of LSBs, and its eventual link to the one of UDGs, is largely undetermined and only marginally studied.
Previous  theoretical works agreeably suggest that  high angular momentum haloes naturally form LSB discs, highlighting  the importance of halo spin in the creation of LSBs (\citealt{Dalcanton97,dutton07}, and references therein). 
Understanding how the broad population of  LSBs form and evolve,  how they are linked to their DM haloes, and, most importantly, how  they fit within the current cosmological model of galaxy formation, is an issue that must be addressed in order to explain the ever increasing number of LSBs observed in our Universe.

In this paper, we make use of state-of-the-art, sophisticated hydrodynamical cosmological simulations,  to answer these questions, pointing out the important role played by angular momentum in creating LSB galaxies. We select a sample of  simulated galaxies within a narrow range of stellar masses but with a wide range of surface brightnesses. The main goal of this work is to find out \textit{what causes the  variation in effective radius (or surface brightness) at a fixed stellar mass}, or else, why some galaxies  end up being lower surface brightness than other galaxies of similar \mstar.

The manuscript is structured as follows: in \Sec{sec:sims} we describe our simulation sample (\cref{sec:nihao}) and the selection criteria of simulated LSB candidates (\cref{sec:selec}), and compare their HI gas content, \mstar\ and \Reff\  with observational datasets in \cref{sec:compare}; in \Sec{sec:propLSB} the properties of simulated LSBs and HSBs are presented, including their surface  brightness profiles and S\'{e}rsic index (\Cref{sec:sbprofile}), photometric images (\Cref{sec:images}), rotation curves (\Cref{sec:vcirc}) and star formation histories (\Cref{sec:sfhsection}); in \Sec{sec:formation}  the formation and emergence of simulated LSBs is studied, addressing the role of DM haloes profiles, concentration and spin  (\cref{sec:gamma},\cref{sec:conc-spin}), mass accretion histories (\cref{sec:mergers}), redshift of last major merger  (\cref{sec:mergers_z}), and finally orbital alignment at merger time  and gas alignment at high $z$ (\cref{sec:mergers_orb},\cref{sec:gas_aligned}); in \Sec{sec:obs_pred} we offer some observational predictions which could be  tested once spatially resolved star formation rates (SFR) in LSB would become accessible; finally in \Sec{sec:conclusions} we conclude our manuscript specifying the main findings on the formation of LSBs, and  highlighting similarities and differences  with respect to  previously studied UDGs.

\begin{figure*}
    \hspace{-0.6cm}
    \includegraphics[width=6cm]{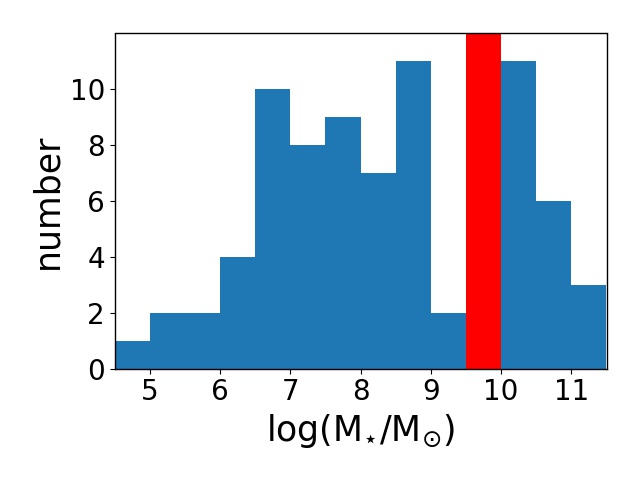}
    \includegraphics[width=6cm]{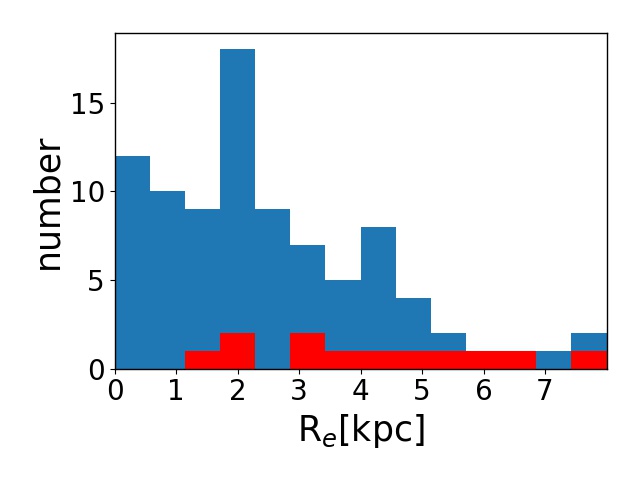}   
    \includegraphics[width=6cm]{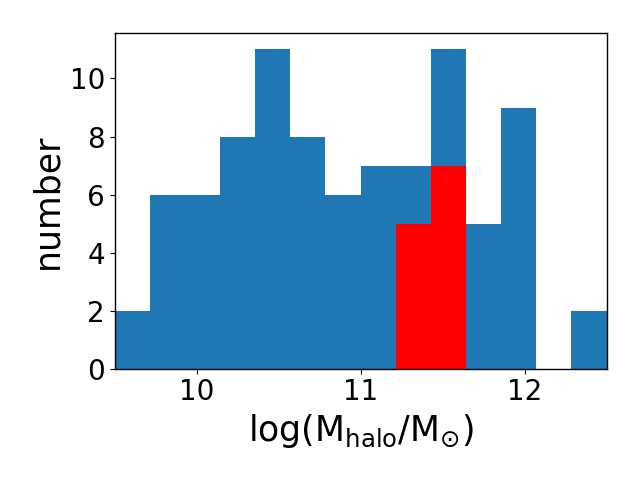}
    \caption{Selection criteria of our sample, compared to the full NIHAO galaxies population (blue histograms). From left to right, we show the stellar mass, effective radius and halo mass of the 12 selected galaxies, as red histograms. Galaxies in a very narrow \mstar\ range exhibit a large variation in their \Reff.}
    \label{fig:selection}
\end{figure*}

\section{Simulation sample} \label{sec:sims}
\subsection{The NIHAO simulations} \label{sec:nihao}
The Numerical Investigation
of a Hundred Astrophysical Objects (NIHAO) project \citep{Wang15}, is a series of 125 cosmologically simulated, zoom-in galaxies,  evolved using  the SPH code Gasoline2.0 \citep{Wadsley17}. The code includes a subgrid model for turbulent mixing of metals and energy \citep{wadsley08}, ultraviolet  heating, ionization and metal cooling \citep{shen10}.
Star formation and feedback follows the model used in the  Making Galaxies In a Cosmological Context simulations (MaGICC) \citep{stinson13}, that for the first time reproduced galaxy scaling relations over a wide mass range \citep{brook12b}, adopting a threshold for star formation of $n_{\rm th}$$>$$10.3 \rm cm^{-3}$ and a \citet{Chabrier03} IMF.

Stars feed energy back into the ISM via blast-wave supernova  feedback \citep{stinson06} and early stellar feedback from massive stars. Particle masses and force softenings of the NIHAO suite  are chosen to resolve the mass profile to below 1$\%$ of the virial radius at all masses, ensuring that galaxy half-light radii are  well resolved.
The NIHAO galaxies are all centrals and isolated and cover a broad mass range, from dwarfs to Milky Way mass, and represent an unbiased sampling of merger histories, concentrations and spin parameters. All  simulated galaxies lie  on  abundance matching predictions from $z$=0 to $z$=3 \citep{moster13}, having the expected \mstar\ for each \mhalo.
The NIHAO project satisfactorily reproduces  realistic galaxies in terms of their \mstar, SFH, metals and DM distribution \citep[e.g.][]{tollet15,obreja16}.  

The  haloes are identified using the AHF\footnote{http://popia.ft.uam.es/AHF/Download.html} 
halo finder \citep{Knollmann09} and partially analysed with the \textit{pynbody}\footnote{https://pynbody.github.io/pynbody/installation.html} package \citep{Pontzen13}.

\subsection{Selection of LSB candidates} \label{sec:selec}
We start by selecting, within the full sample of NIHAO galaxies, those objects whose \mstar\ falls within the range of observed LSBs.
Since the vast majority of known `classical' LSB  have stellar masses comprised between 10$^9$ and 10$^{10}$\msun,  as reported in \citet{Impey97}, we focus on simulated galaxies within such range.
Moreover, we note that diffuse, dwarf LSBs, with \mstar$\leq$10$^9$\msun (i.e. UDGs) have already been  studied and analysed elsewhere \citep{dicintio17,chan18,Jiang18}.

For  this work, we specifically select 12 galaxies within a narrow stellar mass range, $10^{9.5}$$<$\mstar/$\rm M_{\odot}$<$10^{9.9}$, in order to exclude possible mass-dependent effects on our results.
Although having a very similar \mstar, this set of 12 simulated galaxies show an extreme variation in their effective radii, with the largest object having \Reff$\sim$8 kpc and the less extended ones \Reff$\sim$1.5 kpc.
This is shown explicitly in \Fig{fig:selection}, in which, from left to right, the stellar mass, effective radius and halo mass of our selected sample (red histograms) against the full sample of NIHAO galaxies (blue histograms), is shown.

\begin{figure}
    \includegraphics[width=\columnwidth]{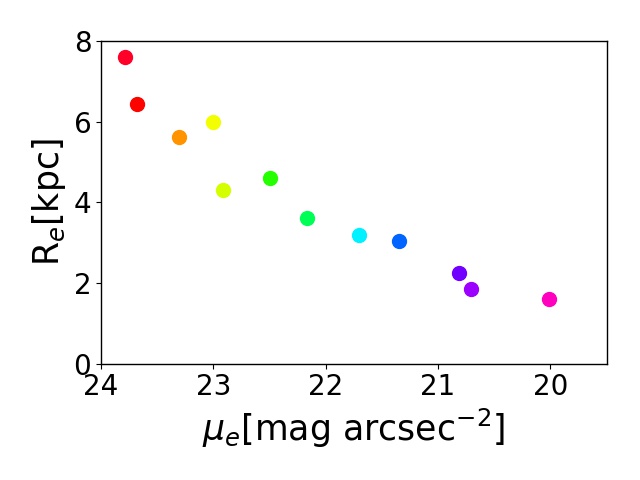}
\caption{Correlation between the effective radius and the effective surface brightness of our sample galaxies. The \SB is further indicated with different colours, from red (LSB) to violet (HSB).}
    \label{fig:reff-sb}
\end{figure}

In \Fig{fig:reff-sb} we demonstrate that the effective radius is a good proxy for the effective surface brightness, given the definition $\mu_e$= L/2$\pi$R$_e^2$, in which L is the luminosity of the galaxy in the desired band. Through this paper, we will work in V-band. No dust attenuation is considered.
\Fig{fig:reff-sb} also demonstrates that several of this suite of simulated galaxies are within the LSB regime.
 Although a strict definition of  minimum \SB required to be classified  as LSB does not exist, we observe that the vast majority of the LSB from the \citet{Impey97} sample has a central surface brightness, in V-band, of 22.0-25.0 mag/arcsec$^2$: we treat our galaxies as a continuum from HSB to LSB, and pay special attention to the lowest surface brightness objects with $\mu_e$$>22$mag/arcsec$^2$.
\begin{figure}
    \includegraphics[width=\columnwidth]{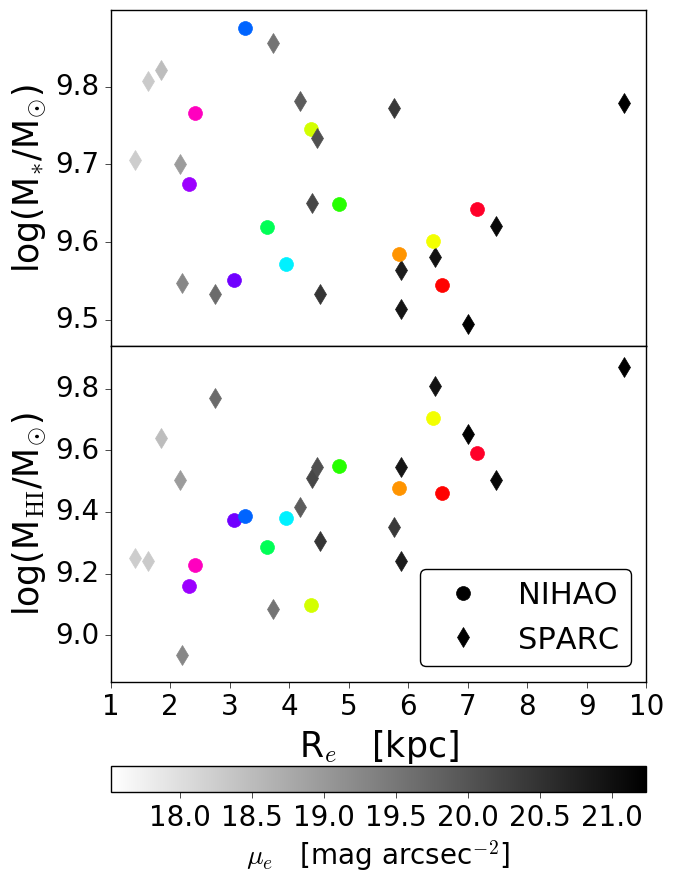} 
\caption{Stellar mass (top panel) and HI gas mass (bottom panel) vs effective radius, for galaxies in the simulated NIHAO (circles)  and observed SPARC (diamonds)  sample, within the same mass range. The effective radius has been calculated in K-band for the NIHAO sample, as a proxy for the 3.6$\mu$m IRAC band of the SPARC data. The colour scheme of NIHAO is the same as in \Fig{fig:reff-sb}, while SPARC data are grey-shaded according to their \SB in 3.6$\mu$m band. The same trend can be observed for both datasets, with NIHAO covering well the full parameter space defined by SPARC galaxies.}
    \label{fig:mstar-sparc}
\end{figure}
\begin{figure}
    \includegraphics[width=\columnwidth]{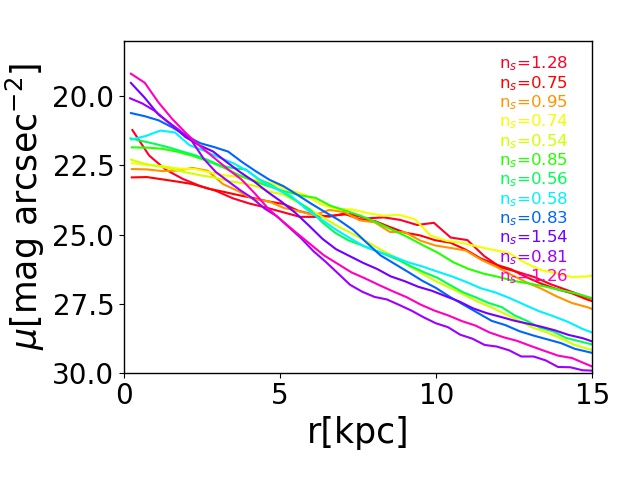}  
\caption{Surface brightness profiles of simulated galaxies, in V-band, and S\'{e}rsic index fit results. The fit has been performed between 0 kpc and 2$R_e$.The shape of \sur profiles for LSBs are similar to what reported in \citet{Pizzella08}.}
    \label{fig:Sersic}
\end{figure}

\begin{figure*}
    \hspace{-.5cm}
    \includegraphics[width=18cm]{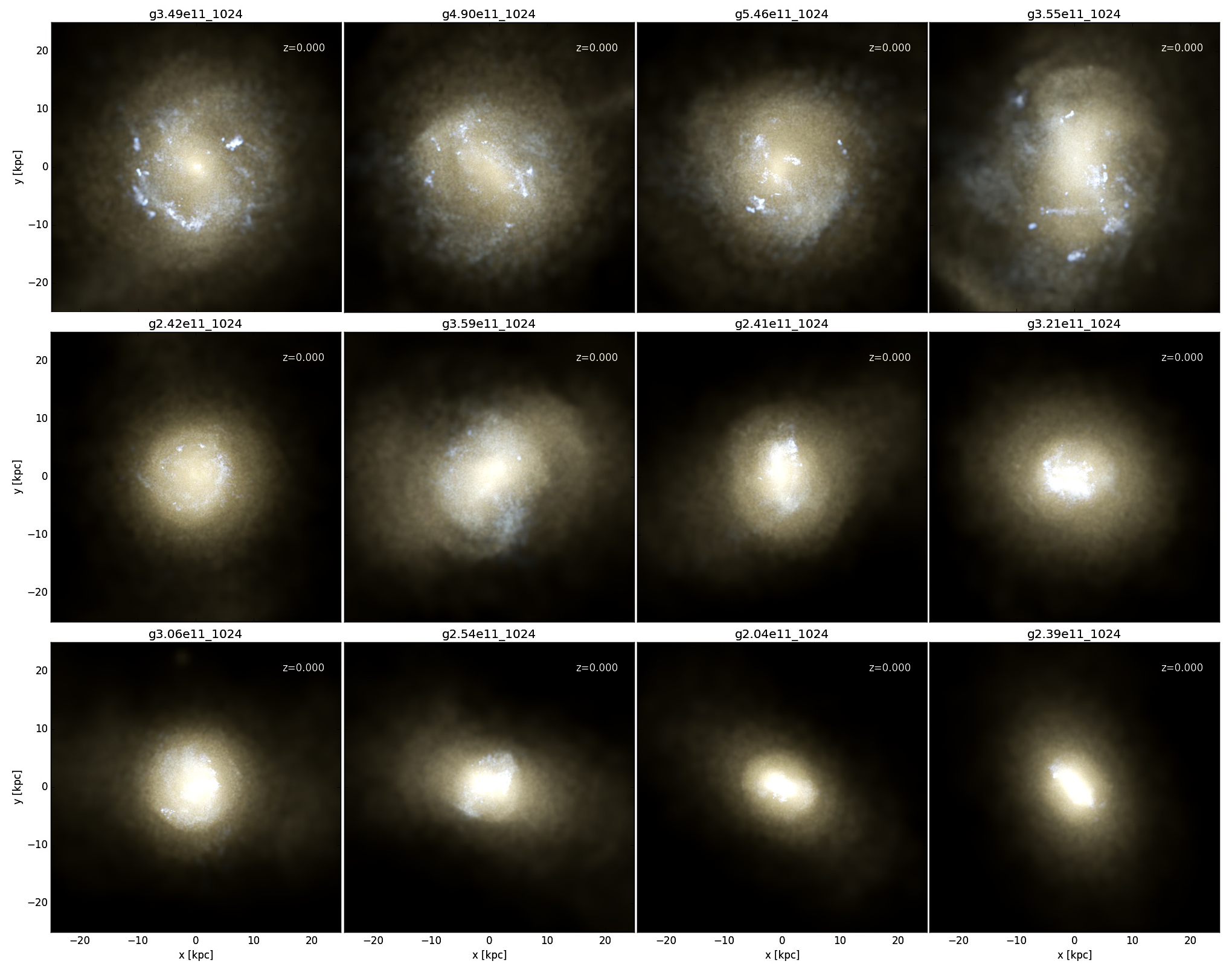}    
    \caption{Multi-band rendering of the stellar light emitted by the sample NIHAO galaxies, down to a surface brightness of $\sim$32 mag/arcsec$^2$. The galaxies are shown face-on and are ordered, from top-left to bottom-right, by their $\mu_e$.}
    \label{fig:rendering}
\end{figure*}

\subsection{Comparison with observational data} \label{sec:compare}
We firstly check that our simulated galaxies match the properties of the observed ones within a similar mass range. To this aim, we compare to galaxies form the SPARC  (Spitzer Photometry and Accurate RCs) data set \citep{lelli16}.
SPARC is a sample of 175 nearby galaxies with homogeneous
Spitzer photometry at 3.6$\mu$m and high-quality rotation curve data from previous HI/H$\alpha$ studies. Though not a complete sample, it is representative of  disc galaxies spanning a very wide range in surface brightnesses and  luminosities, therefore ideal for our comparison.

We select observed galaxies within the same \mstar\ range as our selected simulated galaxies: within this mass range, the SPARC  dataset contains both high and low-surface brightness objects, just like our simulations do.
We show a comparison of NIHAO and SPARC data in \Fig{fig:mstar-sparc}:  the top panel shows \mstar\ vs \Reff\   while in the bottom  M$_{\rm HI}$ vs \Reff\ is represented.  
Amongst our available bands for analysing the simulations, the K-band is the closest match to the 3.6$\mu$m Spitzer-IRAC: we therefore use this band to compare the effective radii of NIHAO and  SPARC datasets.
 NIHAO galaxies are shown as circles and colour-coded  as in \Fig{fig:reff-sb}, based on \SB in V-band\footnote{We verified  that the \SB computed in K-band for NIHAO galaxies provides similar values as the ones observed in SPARC, i.e.  18-22 mag/arcsec$^2$, and we showed this in Appendix A.}, while SPARC data are shown as diamonds and grey-shaded according to their \SB at 3.6$\mu$m.
 For SPARC, the stellar masses are computed assuming a constant stellar mass-to-light ratio of M/L$_{[3.6]}$=0.5, which is consistent with a \citet{Chabrier03} initial mass function. In NIHAO, the  neutral hydrogen mass fraction is computed using the post-processing methods from \citet{rahmati13b} that account for the effects of self-shielding and radiation from local star forming regions (see \citealt{gutcke16} for  details).
 
 A very good match between the observational and the simulated dataset can be appreciated in \Fig{fig:mstar-sparc}. The  overlap in \Reff, \mstar, M$_{\rm HI}$  and \SB is quite remarkable, despite the narrow mass range selected:  galaxies with a low \SB have on average larger HI gas fractions, the sizes covered by our simulated NIHAO sample is in close agreement with the reported sizes of SPARC objects, and both sample span a range of $\sim$4 order of magnitudes in $\mu_e$, from LSB to HSB.
 Interestingly, in this  stellar mass range NIHAO can reproduce very well the full observed range of sizes, producing both compact as well as large objects, unlike in lower \mstar\ ranges, where most of the present day simulations preferentially reproduce diffuse, large galaxies \citep[e.g.][]{dicintio17,chan18,Jiang18}, but struggle at making more compact objects. 
 Given the similar trends observed  in SPARC and NIHAO within the range \mstar=10$^{9.5-9.9}$\msun, we are confident that we are using a suitable simulation set for our study.

\noindent
\section{Properties of the sample} \label{sec:propLSB}
Some relevant properties of the 12 selected simulated galaxies are presented in this section.

\subsection{Surface brightness profiles} \label{sec:sbprofile}
The surface brightness profiles of the simulated galaxies are explicitly shown in \Fig{fig:Sersic}, together with the results of performing a S\'{e}rsic fit between 0 and 2\Reff. As in previous figures, the galaxies are colour coded by their \SB at $z$=0.
We firstly note that LSB galaxies in our sample tend to have a relatively flat central surface brightness profile, with a central value as low as 23 mag/arcsec$^2$, while higher surface brightness objects show a peak in their central $\mu$, indicative of the presence of a bulge. 
An exception to this is represented by the lowest $\mu$ galaxy, shown in red, which has an increase in its $\mu$ within the inner $\sim$ 2 kpc indicating that simulated LSB galaxies can form small bulges (the central $\mu$ is nevertheless lower than any HSB object in our sample).

The extent of the stellar discs can also be derived from this figure: at a fixed $\mu$ of 26 mag/arcsec$^2$, for instance, LSBs reach a radius $\geq$10 kpc, while the stars of HSB galaxies are confined within 6-8 kpc.
The S\'{e}rsic index n$_s$ of LSBs is, on average, less than one, while HSBs  have a mean of n$_s$$\geq$1.

\subsection{Photometric images} \label{sec:images}

\Fig{fig:rendering} shows  IVU-bands images of stars at $z$=0, for each 
galaxy inclined face-on, and ordered from the lowest surface brightness (top left) to the highest  \SB one (bottom right). 
What appears immediately clear is the morphological dependence on surface brightness: galaxies with a low \SB show more extended stellar discs, small or no central bulges, and on-going star formation in the outskirts of the disc, while galaxies with increasingly high \SB have less extended discs, a conspicuous central stellar bulge, and recent star formation mostly confined within the central few kpc.

\begin{figure*}
  \includegraphics[width=17.5cm]{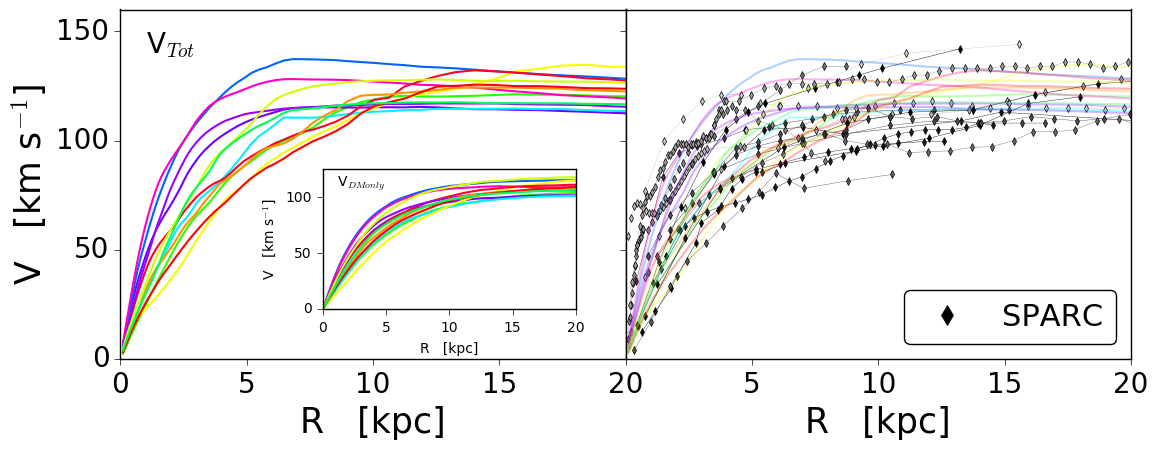}
    \caption{\textit{Left panel:} Rotation velocity of  simulated NIHAO galaxies in the  mass range 9.5<log(\mstar/\msun)<9.9. The large figure shows the total rotation curve, while the insert shows the contribution of the DM halo only. LSB galaxies have a shallow, slowly rising rotation curve, compared to higher surface brightness objects in the same mass range. LSB are DM dominated within their inner radii, while HSBs show a considerable baryonic component at their center. \textit{Right panel:} Comparison between observed and simulated  total V$_{\rm circ}$. The observational data are taken from the SPARC  dataset \citep{lelli16}, selecting galaxies within the same mass range as the NIHAO sample, and shading them by $\mu_e$, from low-(black) to high-(light grey) surface brightness, as in \Fig{fig:mstar-sparc}. The same trend as in the simulations is observed, with the LSB showing slowly rising rotation curves and HSBs having centrally steep V$_{\rm circ}$. Overall, there is good agreement between simulations and observations in the shape and amplitude  of the galaxies' rotation curves in the selected mass range.}
    \label{fig:vcirc}
\end{figure*}

\subsection{Rotation curves} \label{sec:vcirc}
In this section we analyze the circular velocities V$_{\rm circ}$ of simulated galaxies within our selected mass range, and compare them with  the HI/H${\alpha}$ rotation velocities from the  SPARC dataset \citep{lelli16}. SPARC RCs have been corrected for beam smearing, inclination and pressure support.  In the observational sample, we apply the same selection cut as in the simulations, with \mstar\ between 10$^{9.5-9.9}$\msun. In \Fig{fig:vcirc}, left panel,  we show the total circular velocity of the 12 selected NIHAO galaxies, while the insert shows their DM-only rotation curves, colour coded by their \SB as in previous figures, and out to a radius of 20 kpc, to highlight any difference in the inner region. As perhaps expected, LSBs show the  most shallow \textit{total} rotation curves, whilst HSB galaxies show clear signs of a centrally compact, baryonic dominated region within their inner radii, compatible with the morphology and surface brightness profiles explored in \Sec{sec:images} and \Sec{sec:sbprofile}. On the contrary, the DM-only V$_{\rm circ}$ are quite similar amongst galaxies with different surface brightnesses: this suggests that  differences in \SB are not to be attributed to  differences in the DM distribution within galaxies, but rather to how the baryons are distributed in these objects (this aspect will be discussed later  in \Sec{sec:gamma}).

 That LSB galaxies generally have a slowly rising rotation velocity has been shown in several observational works \citep[e.g.][]{deBlok96,deBlok01,McGaugh01,Katz17}. In \Fig{fig:vcirc}, right panel, we explicitly show the comparison between observed and simulated rotation curves, indicating observations from the SPARC dataset as diamonds, with each observed rotation curve grey-shaded by its \SB as in \Fig{fig:mstar-sparc}: in black the lowest surface brightness galaxies, in light grey the HSB ones. Just as in the case of the NIHAO simulations, the SPARC sample shows slowly rising rotation curves for LSB objects, while centrally steep velocities for higher surface brightness galaxies (see also \citealt{Santos18}).

\begin{figure*}
  \includegraphics[width=17.5cm,height=10.5cm]{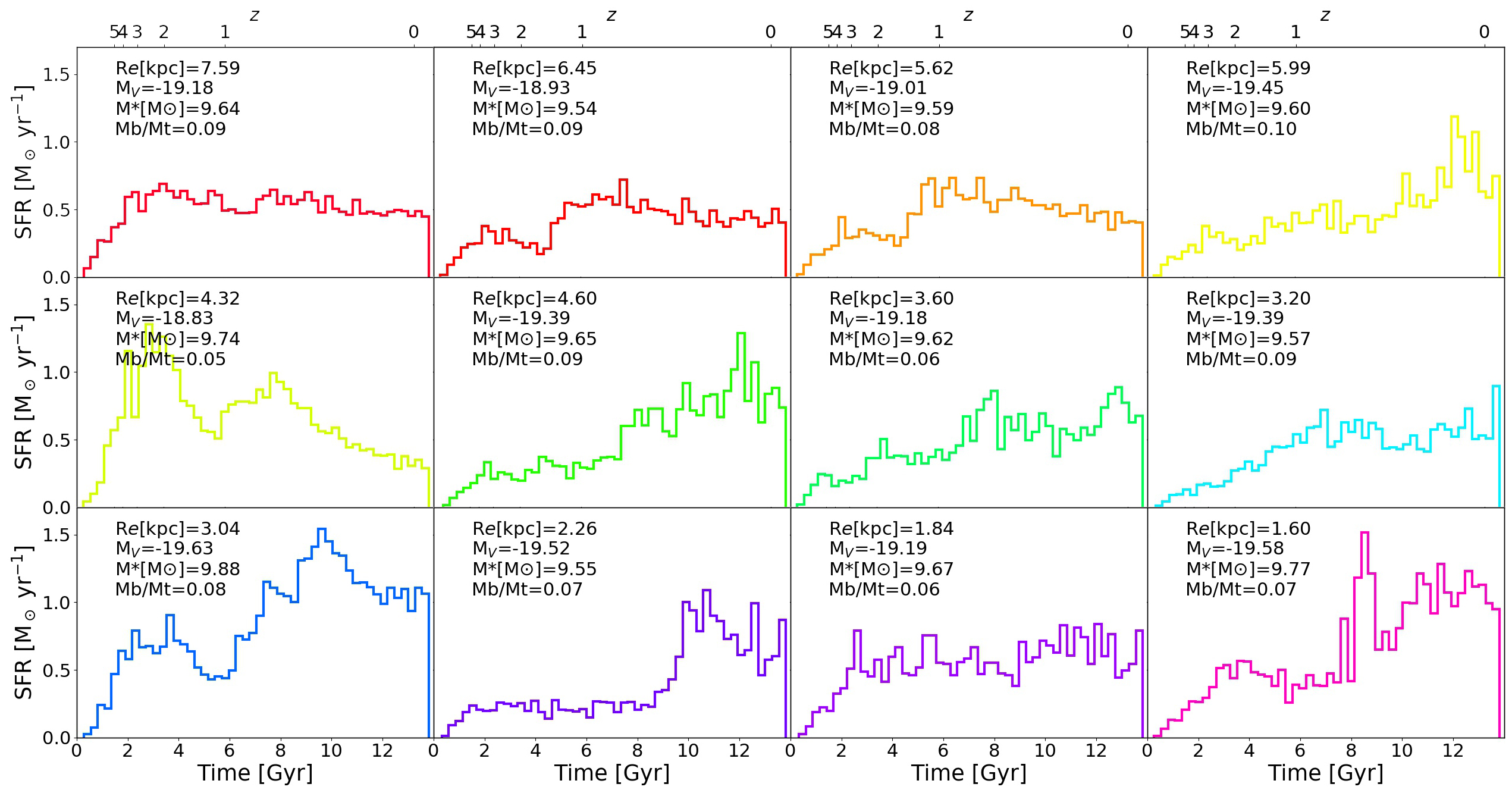}
    \caption{Star formation histories of the sample NIHAO galaxies, ordered  from top-left to bottom-right by their $\mu_e$. In each panel the value of \Reff, magnitude, \mstar and baryon fraction at $z$=0 is indicated. There is no evident correlation between the SFH shape or extent and the size of the galaxy, however it can be noticed how the lowest surface brightness objects in our sample show a continuous, steady SFH all the way to $z$=0, in agreement with observations  \citep[e.g.][]{deBlok96}. The current star formation rate of our simulated LSBs agrees with the observed SFR, as traced by H$\alpha$, of LSBs in a similar stellar mass range, 0.1<SFR<0.4 \msun/yr \citep{mcgaugh17}, and with the NUV-derived SFRs of LSBs in a similar HI mass range, 0.2<SFR<0.8 \msun/yr  \citep{Boissier08}.}
    \label{fig:sfh}
\end{figure*}
\subsection{Star formation histories} \label{sec:sfhsection}
A close look to the SFHs of the simulated galaxies, ordered by their $\mu_e$, is shown in \Fig{fig:sfh}. In each panel we highlight the \Reff, M$_{\rm V}$, stellar mass and baryon fraction at $z$=0. It is interesting to note how the effective surface brightness  does not correlate with the shape or extent of the SFHs: galaxies with very similar SFHs can end up being low or high $\mu$ objects. Similarly, we do not find any evident correlation between \Reff\ (or $\mu_e$) and the  baryon fraction at $z$=0.

This finding is crucial  since it starts unveiling which processes are important in the formation of LSBs: if, as in the model  proposed by \citet{dicintio17} to explain lower \mstar\ UDGs, the formation mechanism of LSBs would be mostly due to SNae feedback driven gas outflows that are able to modify both the central DM density and create extended stellar distribution, we should observe a clear signature of this effect in the SFHs of LSBs vs HSBs, the former having a more bursty and extended SFH than the latter. This signature of feedback-driven core creation has also been  recently reported  in observations of dwarf's SFHs \citep{read18} (see also \citealt{ruizlara18} for an attempt in UDGs). Moreover, larger galaxies should retain a higher baryon fraction than galaxies of similar \mstar, but with lower \Reff, since it is precisely the high  amount of gas the element required to create potential fluctuations.

None of these correlations are observed in our sample of simulated galaxies with \mstar$\sim10^{9.5-9.9}$\msun.

\begin{figure*}
     \includegraphics[width=5.7cm]{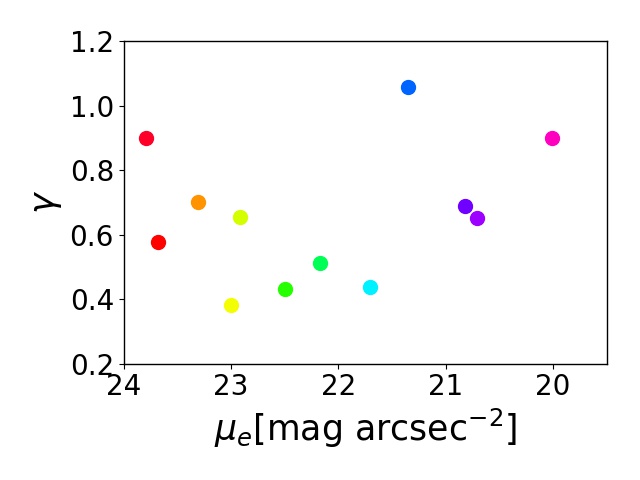}\includegraphics[width=5.7cm]{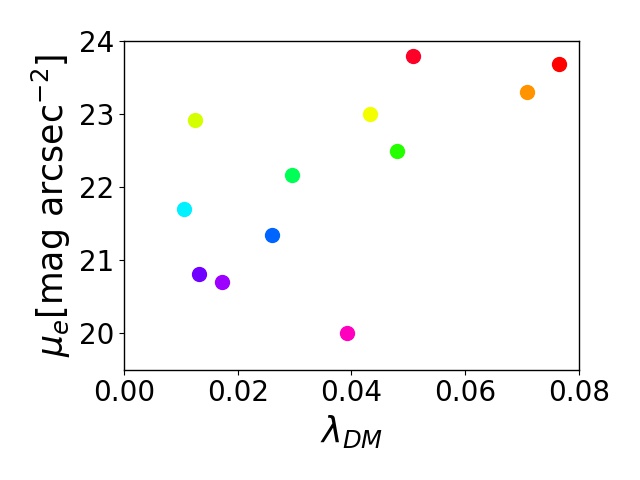}
     \hspace{-0.4cm}
     \includegraphics[width=5.7cm,height=4.2cm]{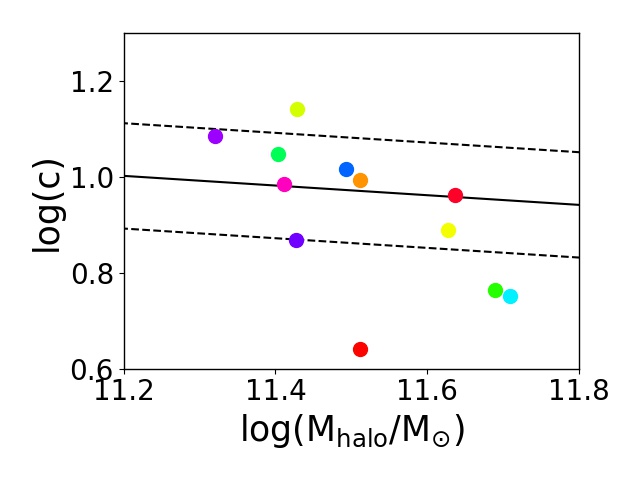}
     \caption{\textit{Left panel:} correlation between the DM inner slope $\gamma$, calculated between 1-2$\%$ Rvir,  and the effective surface brightness of our sample galaxies. No trend is observed between \SB and the central DM slope, the whole sample being in the slightly-expanded regime, with $\overline{\gamma}$$\sim$0.66 for \mstar/\mhalo values of 0.01-0.03 \citep{DiCintio2014a}. \textit{Central  panel:} correlation between  halo spin parameter \citep{bullock01} from  DM-only simulations and  \SB of our sample galaxies. Larger LSBs have larger  DM halo spin, highlighting the role of angular momentum in creating large galaxies. \textit{Right panel:} concentration-mass relation, including 1$\sigma$ scatter, from Planck cosmology  \citep{dutton14}, overimposed over the $c$-$M$ relation of our simulated galaxies.
No relation between \SB and DM haloes concentration is found, both LSBs and HSBs living in similarly average $c$ haloes. Galaxies are colour coded by their $z$=0 $\mu_e$.}
    \label{fig:gamma}
\end{figure*}
\section{The formation mechanism of LSB$\rm_s$} \label{sec:formation}

\subsection{Dark matter profiles} \label{sec:gamma}

Motivated by our previous work on UDGs, and in order to verify whether  LSBs share the same formation mechanism as their dwarfs counterpart, we plot the central DM slope, calculated by fitting a single power law between 1 and 2$\%$ of their respective virial radii, versus effective surface brightness at $z$=0, in \Fig{fig:gamma}, left panel. We did not find any straightforward correlation, the majority of these galaxies having an expanded halo with inner slope of $\gamma\sim$0.4-0.8, in agreement with predictions from \citet{DiCintio2014a,DiCintio2014b,tollet15} for the expected DM profile of galaxies of such stellar mass (or, precisely, of such value of \mstar/\mhalo$\sim$0.01-0.03). Indeed, they lie in the $\gamma$ vs \mstar-\mhalo\ region in which DM haloes are still slightly expanded, but  not at the peak   of core formation, which instead happens for smaller objects of \mstar$\sim$10$^{7-9}$\msun, coinciding with the masses of UDGs.

The lack of a strong correlation between  $\gamma$  and \SB corroborates our  finding of \Sec{sec:sfhsection} that feedback-driven gas outflows are not the primary driver of the formation of galaxies as large as 8 kpc, and thus can not be the only responsible for the emergence of LSBs.

\begin{figure*}
    \hspace{-.5cm} \includegraphics[width=18cm]{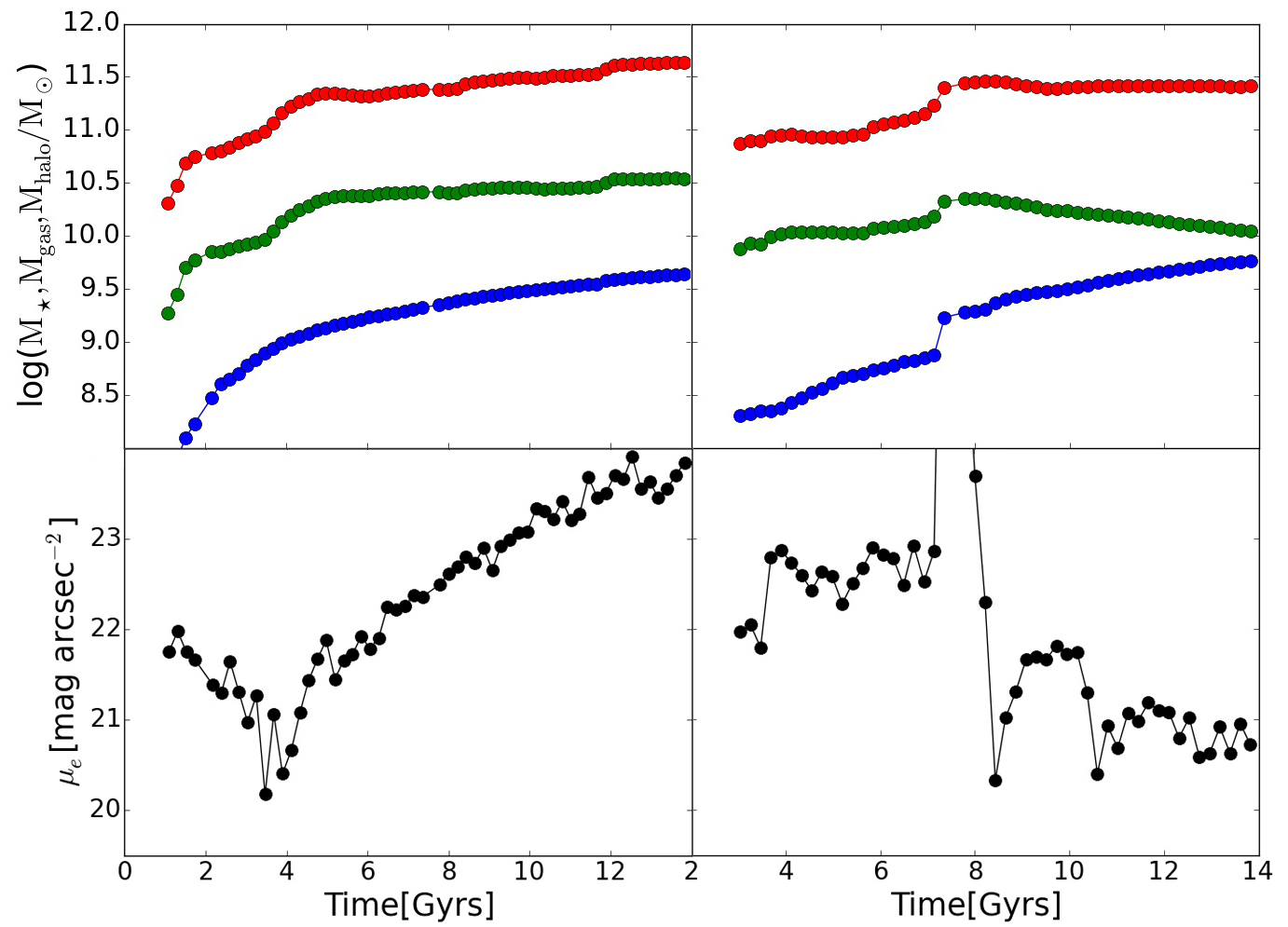}   
\caption{Mass accretion history and effective surface brightness evolution of the lowest (left panels) and highest (right panels) surface brightness galaxies in our sample.
Top panels: Temporal evolution of halo, gas and stellar mass (in red, green and blue, respectively). Bottom panels: \SB vs time. Note how the merger time correlates with a decrease (increase) in \SB for the LSB (HSB) galaxy.}
    \label{fig:evolution}
\end{figure*}

\subsection{Concentration and spin of DM haloes} \label{sec:conc-spin}

We proceed to explore other possibilities in order to explain the emergence of LSBs in our simulations.
One of the formation mechanism suggested for UDGs is the high-spin one, in which UDGs naturally form in the tail of the distribution of  high spin DM haloes \citep{amorisco16}. Although we did not find any correlation between halo spin and sizes in  our simulations when studying small UDG galaxies, we wish to now verify that this finding holds true also for larger LSBs.
In \Fig{fig:gamma}, central panel, we plot the $z$=0 \SB vs DM halo spin, defined as $\lambda=J/\sqrt{2}MVR$ \citep{bullock01}, where $J$ is the total angular momentum of a sphere of radius $R$ containing mass $M$ and with circular velocity $V$: surprisingly, unlike in the case of UDGs, we do find a tight correlation between the final surface brightness of LSBs and the DM halo spin from the corresponding DM only simulation-run. This finding indicates the importance of angular momentum in forming large LSBs, and also highlights once more how UDGs and LSBs seem to have different mechanisms of creation in our simulations.

At last, we verify in \Fig{fig:gamma}, right panel, whether  LSBs tend to form in particularly low concentration haloes. To do so, we  used the DM-only simulated counterpart of our hydrodynamical galaxies, so excluding the effects of baryons on the DM. We did not find any evident  correlation between \SB and halo concentration. In  \Fig{fig:gamma}, right panel, the colour scheme represents the $\mu_e$, just like in previous figures, while the solid and dashed lines are the $c$-$M$ relation in a Planck cosmology and its 1$\sigma$ scatter, respectively \citep{dutton14}. Most of the galaxies studied here lie within the 1$\sigma$ of the $c$-$M$ relation, and no trend is found to support the hypothesis that LSBs live in underdense haloes, compared to HSBs.

\subsection{Evolution of \SB with time and correlation with mass accretion history}  \label{sec:mergers}
We proceed at examining the temporal evolution of the $\mu_e$, together with  the mass accretion history of the lowest and highest surface brightness galaxies in our sample. In the top panels of \Fig{fig:evolution} we show the halo, gas and stellar mass versus time (in red, green and blue respectively), while the effective surface brightness vs time in shown in the bottom panels; the LSB is shown on the left side, while the HSB on the right side.

By inspecting the mass accretion history of the simulated LSB galaxy we can see a clear merger event at t$\sim$3.5 Gyrs, in correspondence of which we observe an unequivocal decrease in surface brightness, moving from a value of 20.3 $\rm mag/arcsec^2$ at t=3.5 Gyrs to $\mu_e$$\sim$22 $\rm mag/arcsec^2$ just after the merger, and finally reaching  $\mu_e$$\sim$24 $\rm mag/arcsec^2$ by $z$=0. In this particular case, the merger ratio between merging  and primary galaxy is of the order of 20$\%$. 
Similarly, we can observe a merger at t$\sim$7 Gyrs for the simulated HSB galaxy, with a merger ratio of about 60$\%$: unlike the LSB case, however, this galaxy starts increasing dramatically its surface brightness just after merger time, moving from an average of $\mu_e$$\sim$22.6 $\rm mag/arcsec^2$ within the first few Gyrs, to $\mu_e$$\sim$21.6 $\rm mag/arcsec^2$ just after the merger, to finally reach the HSB value of $\mu_e$$\sim$20 $\rm mag/arcsec^2$ that we observe today 
 \footnote{Note that the instantaneous decrease in \SB at merger time is an artifact of the halo finder algorithm, that it is unable to differentiate the two galactic discs when they are too close to each other and about to merge.}. 
Note that the larger the merger, the easier it is to identify by the sudden increase in the star formation rate just after merger time, see the corresponding  SFHs in  \Fig{fig:sfh}. The same LSB and HSB galaxies are shown in \Fig{fig:merger_images} in  gas density evolution space.

Mergers thus seem to play a fundamental role in shaping the final surface brightness of galaxies, but while some of them are able to decrease the $\mu_e$, others seem to act in the opposite way, causing an increase in $\mu_e$.
In the HSB case the SF rate is boosted just after the merger, and correspondingly the available gas is consumed, while in the LSB case the SFH and gas consumption proceed more gradually (see also \citealt{SpringelHernquist05}, who found how during some gas-rich mergers not all the gas is necessarily rapidly consumed).
In the next subsection, we aim to identify the mergers parameters and configurations that generate LSBs or HSBs, respectively.

\subsection{The redshift of last major  merger} \label{sec:mergers_z}

\begin{figure}
    \hspace{-.3cm} \includegraphics[width=9cm,height=4cm]{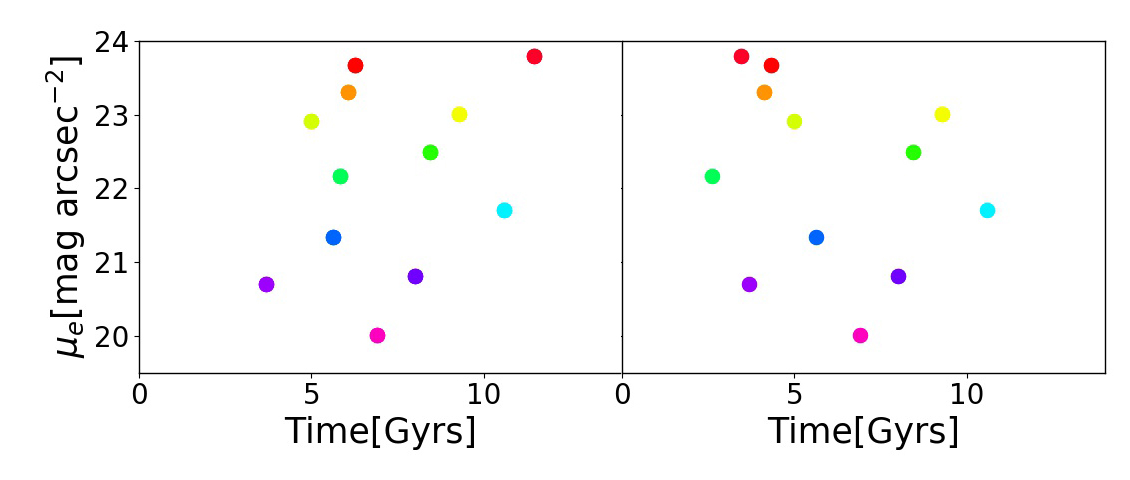}
\caption{Effective \SB vs time of last merger with a merger ratio $\geq$0.1 (left panel) and vs time of largest  merger of each galaxy (right panel). No trend is observed that correlates \SB to the time at which mergers occurred, despite of the size of the merger.}
    \label{fig:merger_time}
\end{figure}

It appears clear, by looking at \Fig{fig:evolution}, that we should  focus on the  mass accretion history of our sample galaxies: the importance of  mergers in producing a variation in \sur seems a crucial point to explore. Not all mergers, however, seem to be associated with a decrease in \sur, as already mentioned in the previous section.
Here, we will explore in detail the role of mergers in shaping the final surface brightness of simulated galaxies.

To this aim, we firstly identify the galaxies' merger histories, by tracking each galaxy back in time: we consider as merging object the most massive structure that was entering within the virial radius of the primary galaxy at a certain time.
We define the merger time as the time at which such structure was lastly found as isolated object (i.e. not yet a satellite of the main halo) by the halo finder algorithm. 
We used different merger ratio thresholds, with a minimum ratio of M$_{\rm merg}/$M$_{\rm prim}$$\geq$0.1, in which M$_{\rm merg}$ and  M$_{\rm prim}$ are the total mass of the merging and of the primary galaxy, at the last snapshot before infall.
Moreover, we only consider mergers that happened  after $z$=2.5, since only a few snapshots are available before that time with the protogalaxy - and eventual merging structures - still containing a limited amount of particles. This is reasonable, considering that the vast majority of stars are born after this time in all cases.

We firstly study the influence of the time of mergers of different sizes in producing higher or lower \sur galaxies.
No correlation between the  redshift of last merger and the \SB at $z$=0 is found, as shown in \Fig{fig:merger_time}, left panel, for a merger ratio of 0.1; we verify that this result holds true for several other  merger ratios, between 0.05 and 0.6, though not shown here.

Additionally, we studied the time of the largest mergers since $z$=2.5, without imposing a fixed merger ratio, such that each galaxy is now associated with its own largest merger with ratio of at least 10$\%$.
Have galaxies with a  low \SB undergone preferentially through a late or early  largest merger?
We found a large variation in the strength of largest mergers, with sizes between 16$\%$ and 75$\%$. In the right panel of \Fig{fig:merger_time} we show the $z$=0 \SB vs time of largest merger for each galaxy: we do not observe  any trend between the two quantities.  
In the next subsection, we  investigate the structural parameters of the mergers.

\begin{figure}
    \includegraphics[width=\columnwidth]{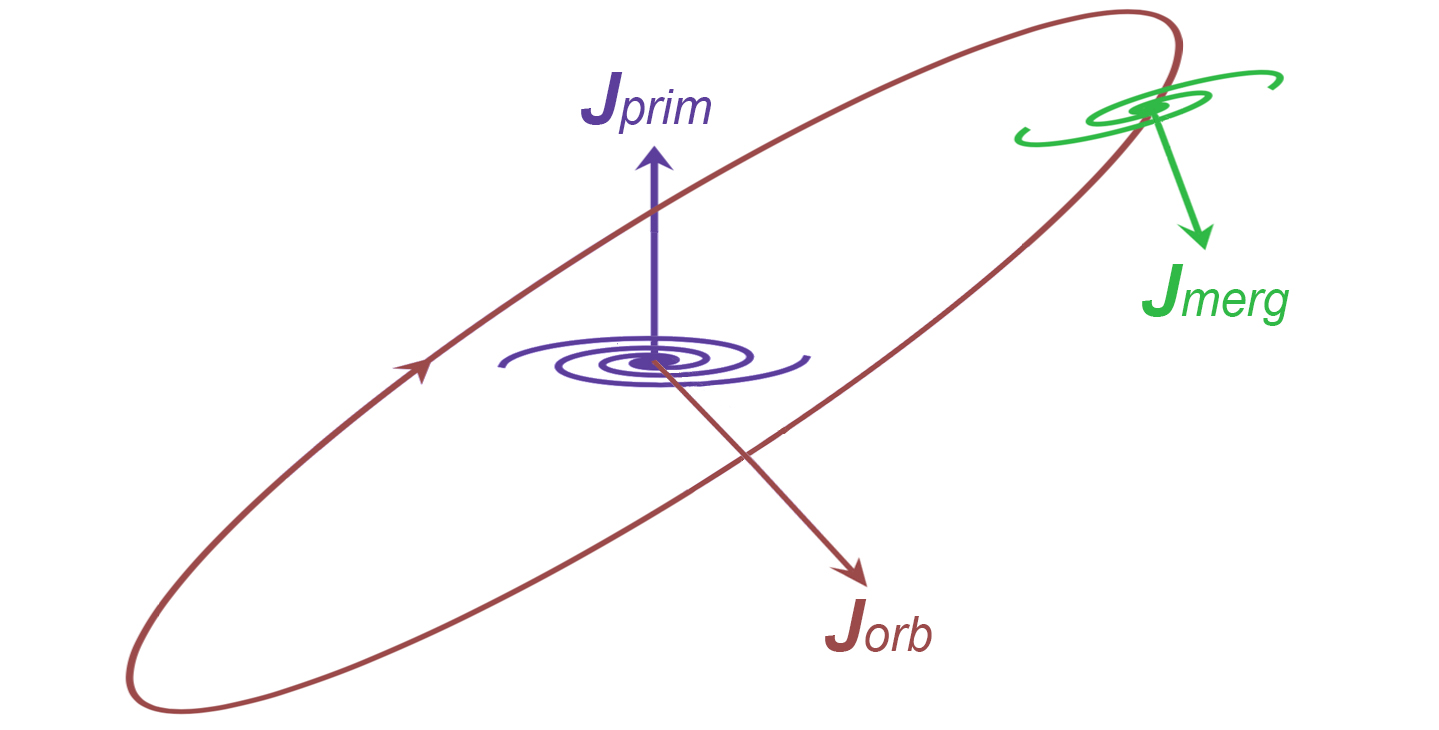}
\caption{Toy model representing the orientation of the angular momentum of the primary and of the merging galaxies, $\Vec{J}_{\rm prim}$ and $\Vec{J}_{\rm merg}$, respectively,  and  of the orbital angular momentum,  $\Vec{J}_{\rm orb}$, at merger time.}
    \label{fig:schematic}
\end{figure}

\subsection{The role of orbital alignment at merger time} \label{sec:mergers_orb}
\begin{figure*}
    \includegraphics[width=18cm,height=4.5cm]{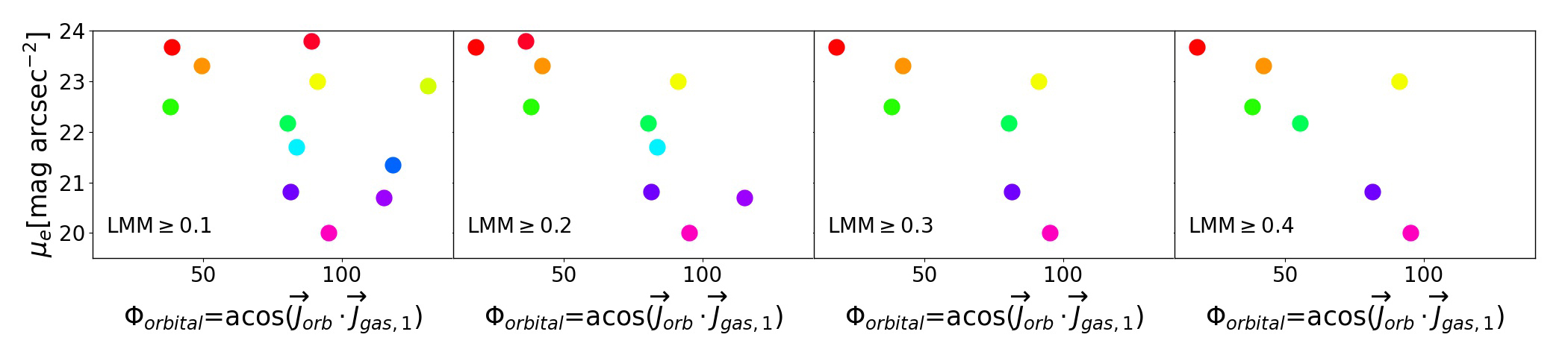}
\caption{Correlation between the $z$=0 effective surface brightness of our sample galaxies and the angle $\phi_{\rm orbital}$, representing the orbital alignment between merging and primary galaxy during merger, for different last major merger ratios (LMM$\geq$10-40$\%$, from left to right). The larger the merger, the stronger the correlation between merger orbital configuration and final $\mu_e$. LSBs have mostly undergone through co-planar, co-rotating mergers ($\phi_{\rm orbital}$$\leq$60$\degree$), while HSBs have experienced  perpendicular mergers (60$\degree$$\leq$$\phi_{\rm orbital}$$\leq$120$\degree$). Colour scheme as in previous figures, based upon each galaxy's final $\mu_e$.}
    \label{fig:merger_angle}
\end{figure*}
Given that the time of mergers does not seem to play a role in determining the final \SB of our galaxies, it is reasonable to assume that some particular merger orientation could be important.
If certain mergers configurations are able to add angular momentum to the primary galaxy and, as a result, decrease its $\mu_e$, we should find a signature of this by studying the orbital configuration  of each primary-merging pair at merger time.

To determine the merger orientation, we define the angle $\phi_{\rm orbital}$  which refers to the alignment of the orbit of the merging galaxy with respect to the primary galaxy,  as: 
\beq
\phi_{\rm orbital}=\rm acos(\overrightarrow{J}_{\rm orb}\cdot\overrightarrow{J}_{\rm gas,prim})
\eeq

Here, $\Vec{J}_{\rm gas,prim}$ is the angular momentum vector of the gas of the primary galaxy, while $\Vec{J}_{\rm orb}=\Vec{r}\times\Vec{v}$  is the angular momentum of the orbit of the merging galaxy relatively to the primary one, with $\Vec{r}$ and $\Vec{v}$ being the position and velocity of the merging galaxy in the reference system of  the main one. The vectors  are all normalized and calculated at the last snapshot before merger.
A schematic view of the orbital angular momentum during merger it is shown in \Fig{fig:schematic}. 

We define the orbital orientation of the merger to be co-planar and co-rotating if the  angle is $0\degree$$<$$\phi_{\rm orbital}$$\leq$$60\degree$ , perpendicular if $60\degree$$<$$\phi_{\rm orbital}$$\leq$$120\degree$ and co-planar counter-rotating if $120\degree$$<$$\phi_{\rm orbital}$$\leq$$180\degree$.

In \Fig{fig:merger_angle} we show the correlation between the final $z$=0 \SB and $\phi_{\rm orbital}$ at merger time, for different merger sizes: from left to right we show, respectively, a last major merger with size ratio of 10-20-30 and 40$\%$. That is, each point represents the orbital alignment at the latest time during which the galaxy underwent a merger of 10-20-30 or 40$\%$. 
While for a low merger ratio ($\sim$10$\%$) we observe little to no  correlation between \SB and $\phi_{\rm orbital}$, the dependence of  final surface brightness from orbital angle at merger time   becomes stronger as we move towards larger merger ratios: already at M$_{\rm merg}/$M$_{\rm prim}$$\geq$0.2 we observe a clear trend such that the lowest \SB galaxies have undergone through a co-planar co-rotating merger ($\phi_{\rm orbital}\leq$60$\degree$), while the highest \SB objects have experienced  a perpendicular merger   (60$\degree\leq\phi_{\rm orbital}\leq$120$\degree$)\footnote{We performed an extra test by defining the angle between the angular momenta of the primary and of the merging galaxy, irrespective of their orbital configuration, but we did not observe any correlation with $\mu_e$. This demonstrate that the alignment of $\Vec{J}_{\rm orb}$  is more important than the relative alignment between $\Vec{J}_{\rm prim}$ and $\Vec{J}_{\rm merg}$.}.\\

This can be understood in terms of angular momentum: galaxies which had co-planar mergers had their total angular momentum increased,  giving rise to large \Reff\ and LSB discs. Oppositely, a perpendicular merger configuration impacts the disc of the primary galaxy in the opposite direction to its rotation, removing  efficiently angular momentum from the existing disc and creating a HSB galaxy. In general, the larger the merger the stronger the correlation with final $\mu_e$.

\begin{figure*}
    \includegraphics[width=18cm]{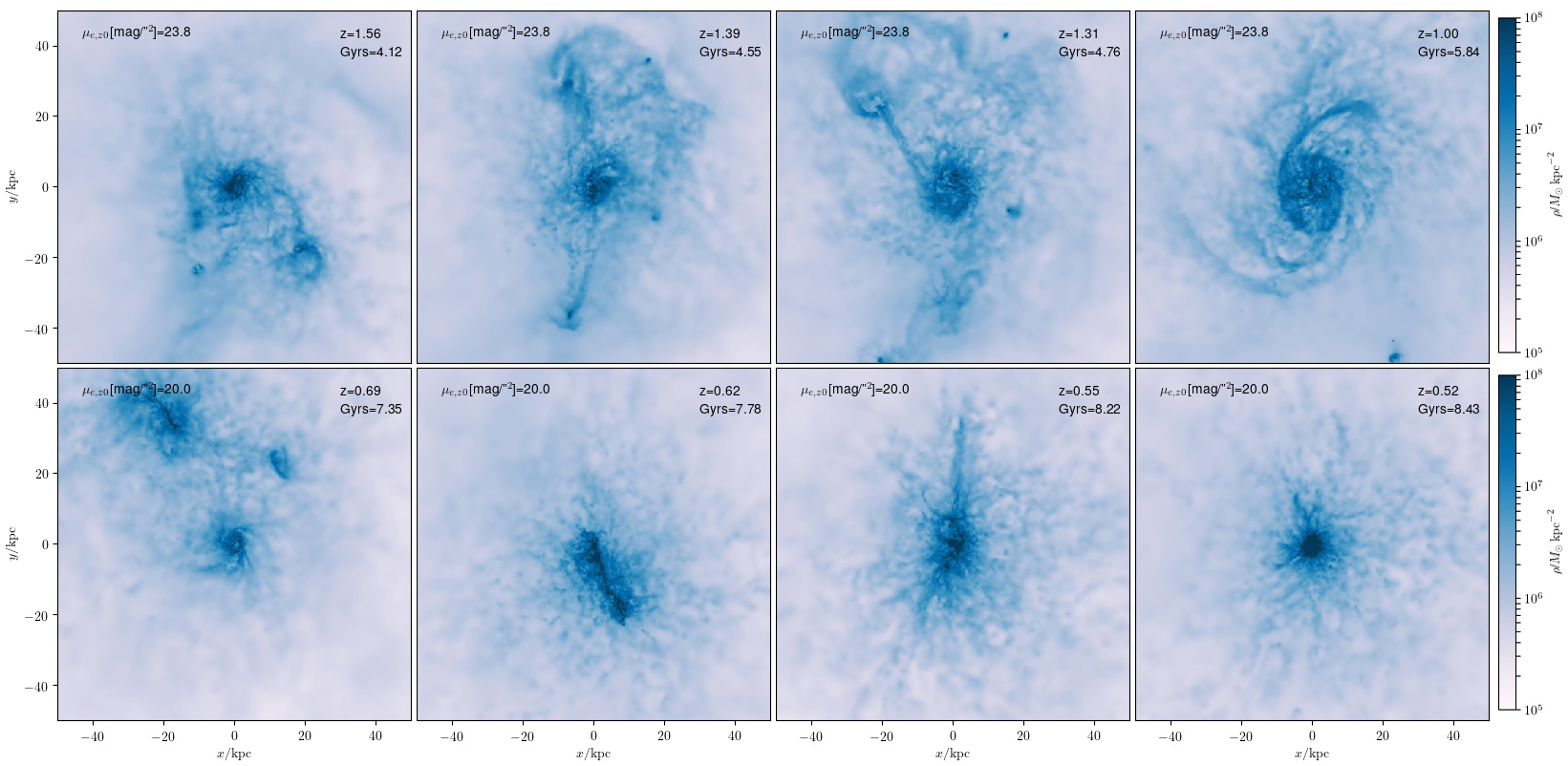}
\caption{Snapshots of  gas density during and after the largest merger of two representative galaxies: the lowest and highest \SB galaxies are shown respectively in the top and bottom rows, with the redshift of each frame highlighted. Both galaxies have a high largest merger ratio ($\geq$20$\%$ and $\geq$60$\%$), thus strongly affecting the dynamic of the central galaxy after the merger. In the LSB case, a co-planar merger can be identified at t$\sim$4 Gyrs, which  adds angular momentum to the disc and whose final product is a large in size, low \SB object (top-right panel). For the HSB galaxy, instead, a perpendicular merger can be observed at t$\sim$7.5 Gyrs: this orbital configuration removes angular momentum from the existing disc and results in a compact, high \SB galaxy just after the merger (bottom-right panel).}    
\label{fig:merger_images}
\end{figure*}

In \Fig{fig:merger_images} we show explicitly time frames of the gas density during and after mergers for the lowest (top row) and highest (bottom row) \SB galaxies in our sample, i.e. the same galaxies analyzed in \Fig{fig:evolution}. A co-planar merger can be identified for the LSB object at t$\sim$4 Gyrs, which efficiently adds angular momentum to the disc and results in a large extended galaxy  with high \Reff\ and low $\mu_e$, just after the merger; the HSB galaxy, instead, shows a clear  perpendicular merger at t$\sim$7.5 Gyrs that gives rise to a compact, low \Reff, high \SB object.  The last temporal frame of \Fig{fig:merger_images}  strikingly shows  the differences in morphology between  the LSB and HSB galaxies, in terms of their gas density and spatial extent, caused by the different orbital configuration during merger.

\begin{figure}
    \includegraphics[width=\columnwidth,height=5.5cm]{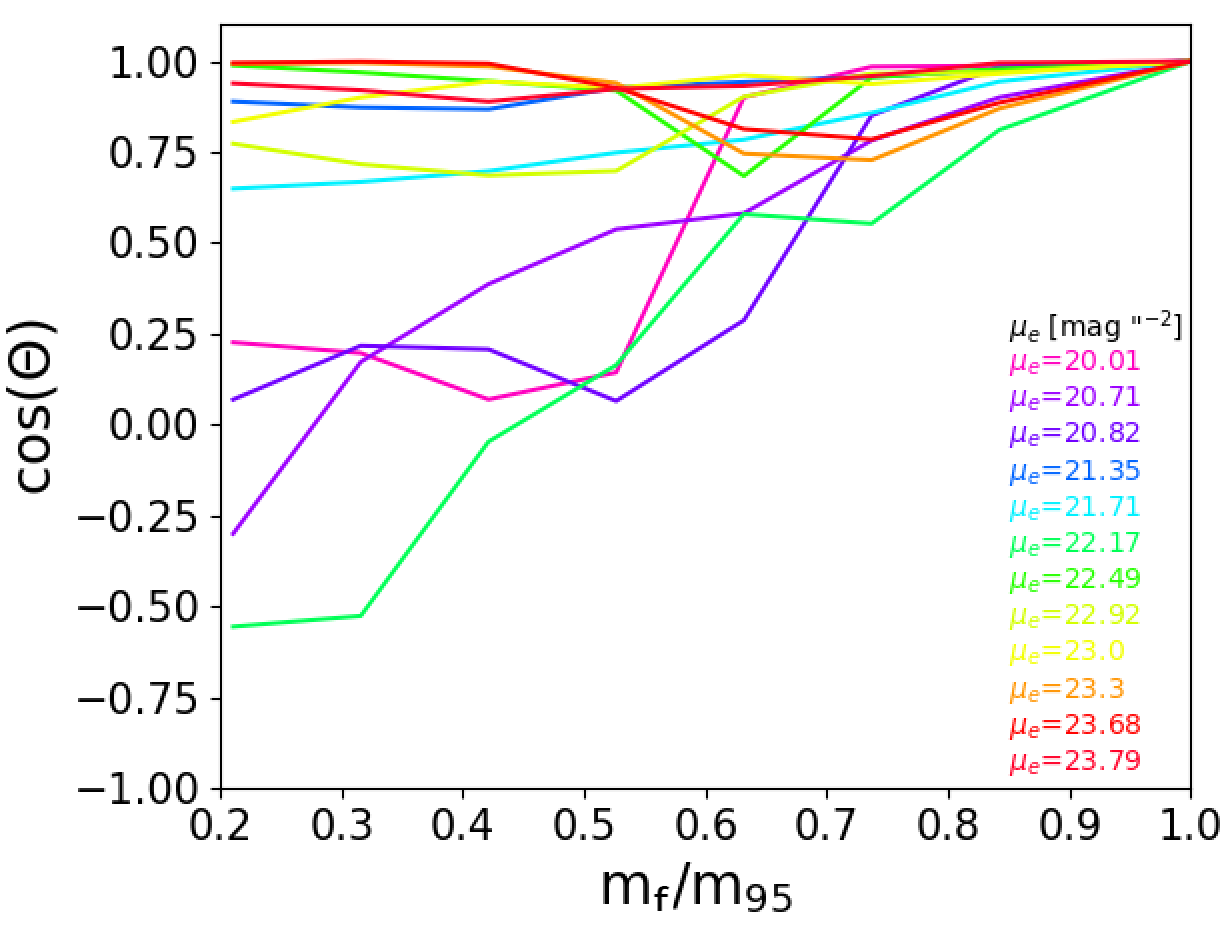}
\caption{Angle between the angular momenta of inner and outer shells  of baryonic material  infalling into the protogalaxy at the galaxy's half-halo mass formation time, vs the ratio mf/m$_{95}$, where m$_{95}$ is the largest shell considered, which includes 95$\%$ of all  baryons that will belong to the galaxy by $z$=0.
For LSB galaxies, inner and outer shells
of baryons, that will
make up the galaxy at 
$z$=0, are well aligned
already at the redshift of half-formation time. On the contrary, HSB galaxies preferentially form from accreting baryons that are misaligned with respect to the central, proto-galaxy.}
    \label{fig:gas_align}
\end{figure}

\subsection{The impact of gas alignment at high $z$} \label{sec:gas_aligned}
Asides from mergers, another factor that could potentially affect the  morphology of simulated  galaxies, and that should be therefore investigated in the study of LSB formation, is the alignment of angular momentum of accreting gas.
As shown in \citet{Sales12}, indeed, the assembly
history, and more precisely the coherent alignment of the angular momentum of baryons
accreting into the galaxy through cosmic time, is a crucial quantity  in determining the final galactic morphology, such that discs form out of gas flowing in with similar angular momentum as the already accreted material, while spheroids form from misaligned gas accretion.

Following the analysis of \citet{Sales12}, in \Fig{fig:gas_align}  we show  the angle $\theta$ between the angular momentum of all the baryons that will eventually end up in the galaxy by $z$=0, and the angular momentum of different interior subsets of such baryons, i.e. those enclosed within spheres containing a given mass fraction (mf) of such baryons: both angular momenta are measured at the time when half of the total halo (dark matter $\texttt{+}$ gas $\texttt{+}$ stars) mass of the galaxy was assembled, $z_{M_{1/2}}$.
So at $z_{M_{1/2}}$, m20 represents the mass fraction of a shell within which 20$\%$ of the baryonic mass is found, whilst m95 represents the mass fraction of a shell that contains 95$\%$ of the  baryonic  mass. 
To this purpose we trace back cold gas and stars from $z$=0 to the half-halo mass accretion time of each galaxy. The x-axis shows mf/m95, with all lines approaching unity as  the enclosed  baryonic mass reaches 95$\%$ of the mass that has been traced. The y-axis shows the cosine of the angle between the angular momentum vector of the baryons within the shell  of mass fraction mf, and those within the shell  of mass fraction m95. 
Lines that are close to unity for every value of  mf/m95, are those galaxies whose inner and outer shells of  accreting baryonic material have a well  aligned angular momentum, at half-halo mass formation time. Galaxies whose accreting baryons are instead misaligned, compared to the central shells, will show up as cos($\theta$) values $<$ 1. The colour scheme is the same used in all previous figures, indicating the \SB of each galaxy.

A clear trend of LSBs having aligned angular momentum for most increasing baryonic shells  can be observed; conversely, the highest \SB galaxies show misaligned accretion of baryons at shells smaller than m95. Galaxies with a halfway values of  \SB show an in-between behaviour.
This analysis demonstrates that asides form mergers, another important   factor  in creating LSB galaxies is the strong alignment of baryonic angular momentum that is infalling into the protogalaxy at high redshift, while misaligned, incoherent  gas accretion at high $z$ results in a subsequent loss of angular momentum, producing HSBs.

The relative strength of the two processes, namely angular momentum acquired through merging galaxies and that gained through aligned inflows of gas, is difficult to assess, as some of the inflowing material into the main progenitor at high $z$ could also be part of a merger later on, and since both effects are inevitable in a hierarchical universe. Ultimately, the tidal field around a galaxy should influence both its smooth accretion as well as the orbital angular momentum of accreting galaxies \citep[e.g.][]{Maller02}.

\section{Observational predictions} \label{sec:obs_pred}
\begin{figure}
\includegraphics[width=\columnwidth]{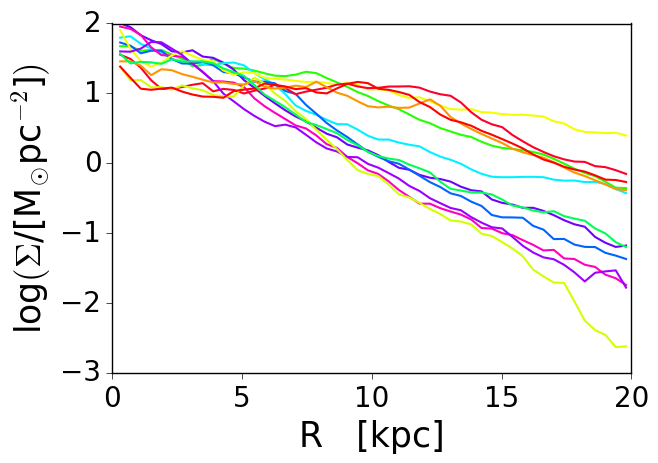}
\caption{HI gas surface density profile for our  simulated galaxies, with the usual colour scheme based on $\mu_e$. While LSBs have a flat HI profile, with densities $\Sigma$$\sim$10\msun/pc$^2$ out to about 13 kpc from their center, HSBs' HI profiles drop below this value already at $\sim$5 kpc. The difference in the HI  profile is reflected in the corresponding radial SFH, as shown in \Fig{fig:2d_sfh}.}
    \label{fig:HI_profile}
\end{figure}

As  shown in \Fig{fig:mstar-sparc}, LSB galaxies tend to have a larger HI gas fraction compared to HSB in the same mass range. But how is this gas distributed across the galaxy?
We show this in \Fig{fig:HI_profile}, 
by plotting the projected HI gas surface density profile of our NIHAO sample, with the usual colour scheme: 
LSB have more extended and flat HI gas profile, with densities of $\Sigma$$\sim$10\msun/pc$^2$ all the way to 13-15 kpc, while in  HBSs the neutral gas falls below such density already at 4-5 kpc. 
Presumably, these differences will produce different signatures in the SFHs of low- vs high-surface brightness galaxies.
Indeed, while the integrated SFHs of LSB and HSB look quite similar (\Fig{fig:sfh}), with  mergers and accretion driving new episodes of star formation in both cases,  it is reasonable to believe that there will be differences in the spatial location at which  new stars are formed,   given the different profiles of  neutral gas in  LSB/HBS, as  in \Fig{fig:HI_profile}.

\begin{figure*}
\hspace{-.2cm}\includegraphics[width=18cm]{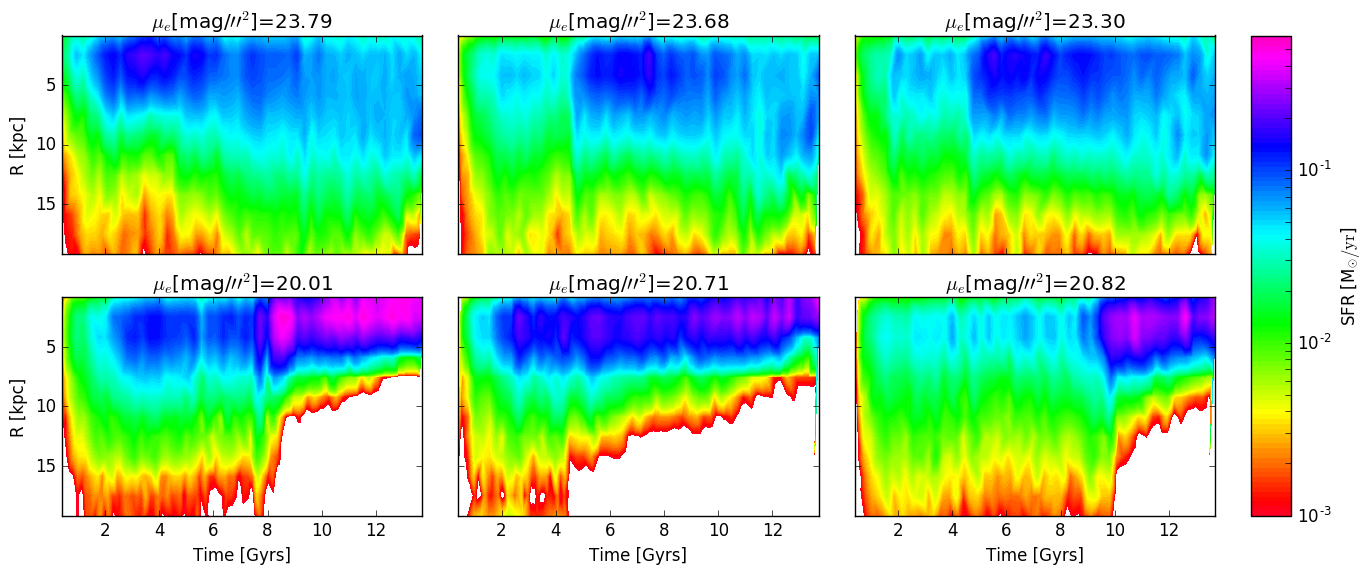}
\caption{2D histograms showing the SFR of simulated galaxies across cosmic time and for  increasing radii from the galactic center.
The top row shows the three lowest \SB galaxies in our sample, while the bottom row indicates the highest \SB ones. The difference between the two groups is quite striking for t>8 Gyrs: while HSBs tend to form the majority of their new stars within the inner 5 kpc, with  no young stars found at radii larger than $\sim$ 10 kpc, in LSBs the star formation  of stars  younger than 5 Gyrs  is spread through the disc, with  young stars found all the way out to 20 kpc. Although the integrated SFH is similar for the two groups (see \Fig{fig:sfh}), the  distribution of young stars within the discs diverges dramatically,  a consequence of   the distinct merger configurations and gas alignment  that give rise to different HI gas density profiles in LSB vs HSB, the former showing the most extended and flat neutral gas profile.}
    \label{fig:2d_sfh}
\end{figure*}

We show this explicitly in \Fig{fig:2d_sfh}, by displaying a 2D histogram of the radial-SFH as a function of cosmic time, for the three lowest (top panels) and three highest (lower panels)  surface brightness galaxies in our sample.
Different colours represents different SF rates, out to a limiting radius of $\sim$20 kpc.
HSBs tend to form most of their new, young stars (age$<$4-5 Gyrs), in a spatial region confined within the inner 5 kpc from their galactic center, a result of perpendicular merger configuration and misaligned gas accretion, that give rise to a central high gas density, leading to  localized SF. LSB, on the contrary, with their flat HI gas density profile translating into higher gas densities at larger radii, resulting from the high angular momentum acquired during co-planar mergers and aligned gas accretion,  are able to sustain the formation of new stars  through  the disc, out to $\sim$20 kpc. 

At the present time, obtaining spatially resolved SFR in such LSBs is still beyond current capabilities. Indeed, SFRs in LSB galaxies have been derived from models that take into account the photometric and chemical evolution of the galaxies \citep[e.g.][]{vandenHoek00}, or by measuring their H$\alpha$ lines \citep[e.g.][]{mcgaugh17}, or by adopting standard calibrations to convert the UV light into
SFRs \citep[e.g.][]{Boissier08}: these techniques provide the mean or present-day SFR in LSBs, without any information about the radial dependence of this quantity.

 In the future, however, new observations and improved technologies will hopefully make possible the derivation of detailed radial SFRs for  these galaxies, validating our prediction.

\section{Conclusions} \label{sec:conclusions}
We explore the formation of Low Surface Brightness (LSB) galaxies using cosmological hydrodynamical simulations from the NIHAO project \citep{Wang15,dutton16}.
For the first time, we show that simulations can reproduce LSBs in the range \mstar$\sim10^{9.5-10}$\msun, coinciding with the galaxy mass where the majority of  LSBs are found \citep{Impey97,Bothun97}: the properties of simulated and observed LSBs agree remarkably. 
The main results of this work are as follows:

\begin{itemize}
\item simulated LSBs are neutral hydrogen rich, log$_{10}$(M$_{\rm HI}$/\msun)$\gtrsim$9.4, have large effective radii, \Reff$\gtrsim$4 kpc, low S\'{e}rsic index, n$_e$$<$1, and slowly rising rotation curves, V$_{\rm 2.5 kpc}$$\sim$50-70 kpc, compared to High Surface Brightness (HSB) galaxies within the same mass range (\Cref{fig:reff-sb,fig:mstar-sparc,fig:Sersic,fig:vcirc});
\item they show an overall extended, continuous, steady, global SFHs,
with current star formation rates 0.2$\lesssim$SFR$\lesssim$0.7 \msun/yr (\Cref{fig:sfh});
\item  they form preferentially in high spin dark matter (DM) haloes, $\rm\lambda_{DM}$$\geq$0.04,  with average concentration parameters following the $c$-$M$ relation (\Cref{fig:gamma}) ;
\item  they live in expanded DM haloes with inner slopes $\gamma$$\sim$0.4-0.6, as a result of  baryonic outflows, although no relation between surface brightness and DM slope is observed in this mass range (\Cref{fig:gamma});
\item they present flat surface brightness, $\mu$, and HI gas density, $\Sigma$, profiles, and a morphological dependence on surface brightness, such that the lowest $\mu$ objects have more extended stellar discs, little or no central bulge, and on-going star formation in the outskirts of their disc (\Cref{fig:Sersic,fig:HI_profile,fig:rendering});
\item \textit{LSBs form through a combination of i) co-planar, co-rotating mergers and ii) aligned infall of gas at early times}.

- We showed the importance of merger parameters in determining galaxy surface brightness by studying the correlation of the angle $\phi_{\rm orbital}$, between the orbital angular momentum vector and the angular momentum of the primary galaxy at merger time, and the final $\mu$ of each galaxy: LSBs have undergone through co-planar mergers, which are able to add angular momentum to the disc and result in  low $\mu$, while HSBs have formed as a result of perpendicular mergers, which efficiently remove angular momentum from the existing galaxy (\Cref{fig:merger_angle,fig:merger_images}); the larger is the merger, the strongest is the correlation between orbital configuration at merger time and final $\mu$, while the specific time at which such mergers occur does not seem to play a role (\Cref{fig:merger_time}); 

- We showed that the alignment of baryons at the time when the galaxy was half its total mass also affects the final surface brightness, by deriving the angle $\theta$ between the angular momentum of inner and outer shells
of baryonic material infalling into the protogalaxy at half-halo mass
formation time, $z_{M_{1/2}}$: for LSBs, the inflowing gas that will compose the galaxy by $z$=0 is well aligned with the protogalaxy already at $z_{M_{1/2}}$, while HSBs preferentially form from accreting baryons that are misaligned with respect to the central object (\Cref{fig:gas_align});
\item their young stellar population (t$_{\rm age}$$\leq$4-5Gyrs) forms throughout the disc, from the inner regions to the outskirts, a result of the flat HI gas surface density extending to large radii (\Cref{fig:HI_profile,fig:2d_sfh}).  The formation of young stars in LSBs proceed at a similar rate  of SFR$\sim$0.1 \msun/yr out to 10 Kpc in the disc, and at about  $\sim$0.01 \msun/yr out to 15 kpc, unlike HSB galaxies whose bulk of young stars is formed solely within the inner $\sim$ 5 kpc from the center, at a higher SFR (\Cref{fig:2d_sfh}).
\end{itemize}

The latter prediction is currently not testable with the present knowledge and  facilities, indeed obtaining spatially resolved SFR in such low \SB systems is still inaccessible. Should deeper observations and better methodologies be available in the future, that will make it feasible the validation of our prediction.
Further, detailed predictions can be offered regarding the rotation vs dispersion support and axis-ratio of LSBs: this is indeed the focus of a forthcoming paper. Here, we anticipate that LSBs in the studied mass range, \mstar$\sim$10$^{9.5-10}$\msun, tend to be rotationally supported and to have an ellipticity between 0.6$<$$\epsilon$$<$0.8, when measured edge-on, compatibly with being typical  disc galaxies, with  large axis-ratio and no bulge component \citep[e.g.][]{Kautsch06}.

Unlike the  scenario suggested by cosmological simulations for the formation of less massive LSB objects, i.e. Ultra-Diffuse galaxies, in which the expanded stellar distribution arises as the result of powerful stellar feedback-driven gas outflows \citep{dicintio17,chan18}, these more massive, `classical' LSBs, have a high enough  mass to be in the realm where stellar feedback alone is not sufficient to create  such large, extended galaxies \citep{governato12,DiCintio2014a,chan15,tollet15}. 

For \mstar$\gtrsim$$10^9$\msun\ the role of angular momentum becomes  predominant, as we move from a \textit{feedback-based} to a \textit{angular momentum-based} formation scenario. This extends upon  previous theoretical work suggesting the importance of high-angular momentum in the formation of LSBs \citep[e.g.][]{Dalcanton97,dutton07,amorisco16}.
We remark  that the proposed formation scenario for LSBs only applies to isolated galaxies in the selected mass range, \mstar$\sim$10$^{9.5-10}$\msun,  most of which are discs in the \citet{Impey97} sample and in our simulations. Lower mass dwarfs ellipticals and dwarf irregulars, with low $\mu_e$, could  result from a combination of strong feedback and environmental effects, though not studied here.

The present work implies that, although apparently dwarf and disc galaxies in the LSB domain share similar observational properties, their formation mechanism is intrinsically different, a reflection of the mass dependence of feedback and accretion phenomena across galaxy populations.
Tentative signatures of this transition regime can  already be found in observations of gas fractions and SFHs of dwarfs and disc LSBs, respectively. 

Results from  \citet{Schombert01} and \citet{McGaugh97} suggest that star formation in dwarf LSBs cannot evolve as smoothly and uniformly as it happens for LSB discs, perhaps pointing towards an intrinsic difference in the type of star formation within the two groups; moreover, LSBs dwarfs seem to have  higher gas fractions than those of their more massive counterpart.
Both these aspects are  detected  in our simulations, with low mass UDGs having  bursty-like SFHs and high gas fractions, f$_g$, up to 97$\%$ (\citealt{dicintio17}, their Fig.4), while more massive LSBs, from  this work, showing continuous SFHs and lower f$_g$$\sim$50$\%$.
\cite{Schombert06} previously noted that the structural differences between gas-rich dwarfs and disc galaxies could be driven by  kinematics, and  \cite{Wheeler17} subsequently showed that gas-rich dwarfs  below \mstar$\sim$$10^8$\msun\ may form as dispersion-supported stellar systems, only mildly rotating.

Ultimately, the results of our work could indicate that we are facing the turning point between ordered rotating discs to  more pressure-supported dwarfs systems.
The details of this transition will be explored and confirmed in detail in a forthcoming work.

\section*{Acknowledgements}
ADC acknowledges financial support from a Marie-Sk\l{}odowska-Curie Individual Fellowship grant, H2020-MSCA-IF-2016 Grant agreement 748213, DIGESTIVO. CBB thanks MINECO/FEDER grant AYA2015-63810-P and the Ram\'{o}n y Cajal fellowship program. ADC acknowledges fruitful discussions with Julio Navarro, Laura Sales and Patricia S\'{a}nchez-Bl\'{a}zquez. 
This research was carried out on the High Performance Computing resources at New York University Abu Dhabi. Additional computational resources were provided by the  {\sc theo} cluster at MPIA and the {\sc hydra} clusters at Rechenzentrum in Garching.


\bibliographystyle{mnras}
\bibliography{archive}

\appendix
\section{\SB in NIHAO and SPARC}

\begin{figure}
    \includegraphics[width=\columnwidth]{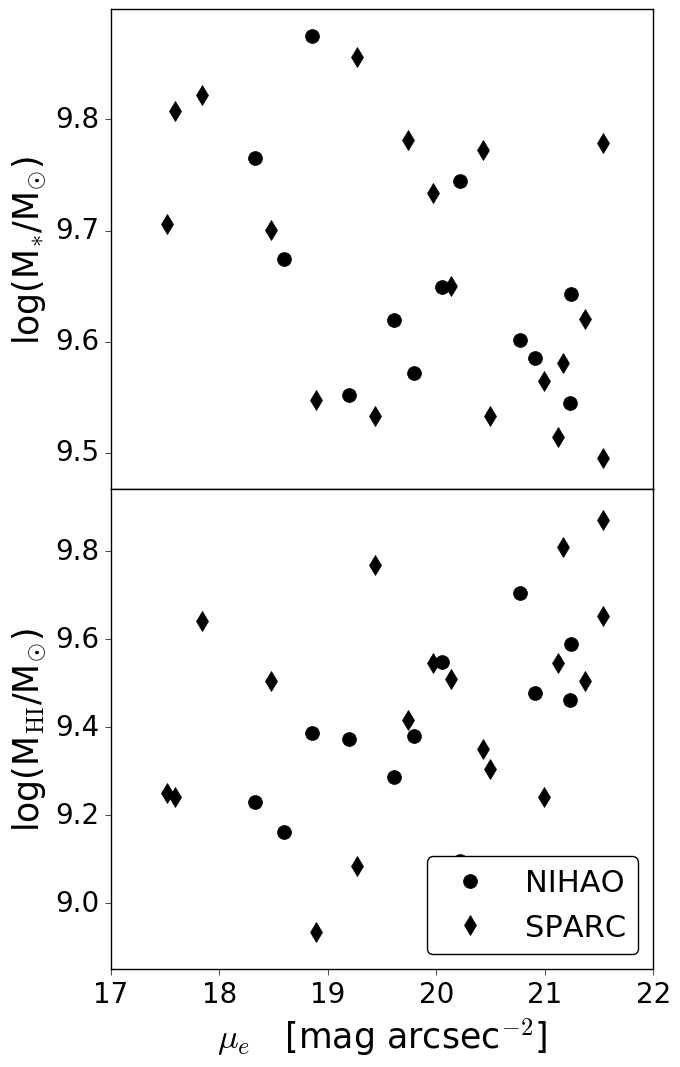}  
\caption{Stellar mass  and HI gas mass  vs effective surface brightness, for galaxies in the simulated NIHAO (circles)  and observed SPARC (diamonds)  sample, within the same mass range. The \SB has been calculated in K-band for the NIHAO sample, as a proxy for the 3.6$\mu$m IRAC band used for the derivation of the SPARC data's \SB.}
    \label{fig:A1}
\end{figure}

We show explicitely the match between the effective surface brightness of our sample of simulated galaxies versus the one of the reference SPARC dataset, in \Fig{fig:A1}.
The K-band is the closest match to the 3.6$\mu$m Spitzer-IRAC band used by SPARC, we therefore use this band to compare the \SB of NIHAO and  SPARC datasets.
 NIHAO galaxies are shown as dark  circles  while SPARC data are shown as dark diamonds. The relation between \mstar\ or \mHI\ and \SB is remarkably similar for observations and simulations, alike what shown already in \Fig{fig:mstar-sparc}.

\bsp	
\label{lastpage}
\end{document}